\documentclass[9pt,twocolumn]{article}

\usepackage{lineno}
\usepackage{amsmath}
\usepackage{amssymb}
\usepackage{braket}
\usepackage{authblk}
\usepackage{graphicx}

\title{Increased instantaneous bandwidth of Rydberg atom electrometry with an optical frequency comb probe}

\author[1,*]{Alexandra B. Artusio-Glimpse}
\author[2,**]{David A. Long}
\author[2,3]{Sean M. Bresler}
\author[1]{Nikunjkumar Prajapati}
\author[1,4]{Dangka Shylla} 
\author[1]{Andrew P. Rotunno}
\author[1]{Matthew T. Simons}
\author[1]{Samuel Berweger}
\author[1,4]{Noah Schlossberger}
\author[2]{Thomas W. LeBrun}
\author[1]{Christopher L. Holloway}

\affil[1]{National Institute of Standards and Technology, Boulder, CO 80305 USA}
\affil[2]{National Institute of Standards and Technology, Gaithersburg, MD 20899 USA}
\affil[3]{University of Maryland, College Park, MD 20742 USA}
\affil[4]{University of Colorado, Boulder CO 80302 USA}

\affil[*]{alexandra.artusio-glimpse@nist.gov}
\affil[**]{david.long@nist.gov}

\begin{document}

\maketitle

\textbf{
We show that the use of an optical frequency comb probe leads to dramatically improved bandwidth (as high as $12\pm1~\mathrm{MHz}$) for the detection of modulated radio frequencies in Rydberg atom-based electrometry.
}

\vspace{1em}
The use of Rydberg atoms for measurements of modulated radiofrequency (RF) electric fields has generated considerable interest due to the combination of wide tunability, low invasiveness, small size, intrinsic accuracy, and traceability to the International System of Units (SI). However, the relatively low bandwidth of these approaches has limited their applications, especially in areas requiring reception of complex waveforms and high data rates. While some recent improvement of the instantaneous bandwidth has been demonstrated through the use of reduced beam sizes~\cite{Bohaichuk2022,Prajapati2022,Hu2023} and strong coupling lasers~\cite{Hu2023,Yang2023_arxiv}, these techniques have drawbacks. Reducing beam sizes to decrease transit times of the atoms traversing the beams necessitates low probe laser power to maintain the intensity. This reduces signal-to-noise ratio, requiring system modifications to maintain reasonable sensitivity levels. On the other hand, a strong coupling laser can drive signal gain at its Rabi frequency without direct loss of sensitivity. However, higher power coupling lasers increase the charge generation at glass surfaces and in the vapor, resulting in electric fields that will distort the Rydberg atom spectra and limit RF field measurements~\cite{Gallagher1988,Ma2020}.

\begin{figure*}[t]
    \centering   
        \includegraphics[width=\textwidth]{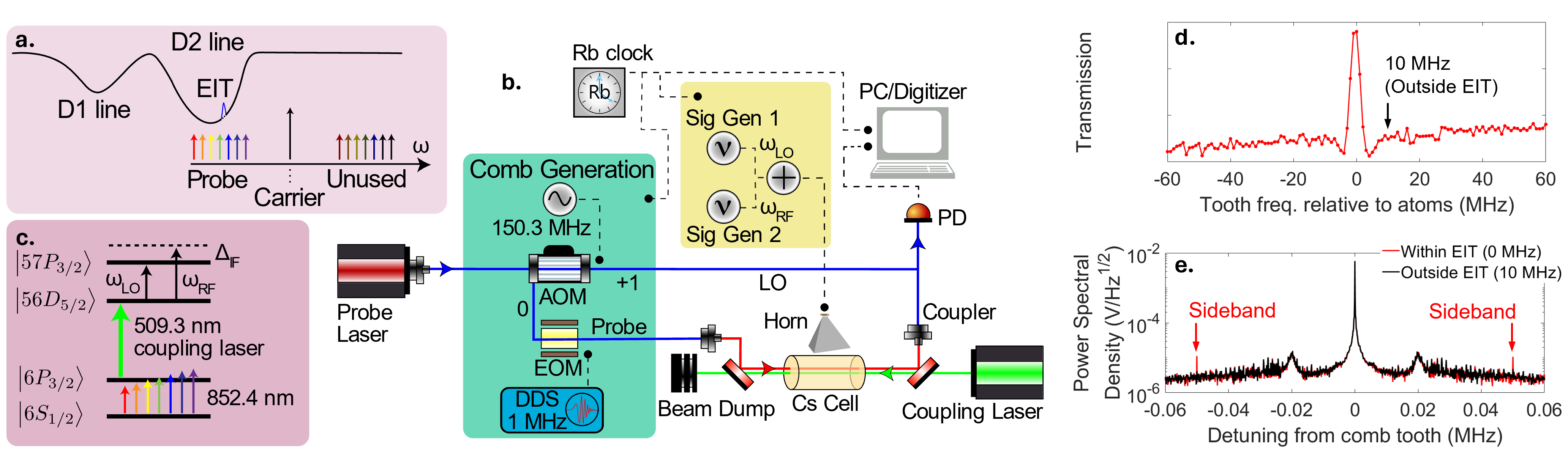}
        \caption{\textbf{a.} Depiction of the $852~\mathrm{nm}$ frequency comb and \textbf{b.} measurement system. The comb includes a strong carrier blue-shifted from the D2 line, a probe comb $200~\mathrm{MHz}$ wide and approximately centered on the $F=4 - F^\prime=5$ transition with $1~\mathrm{MHz}$ tooth spacing, and a second unused comb $1020~\mathrm{MHz}$ above the probe comb. The comb is produced via a chirped waveform from a direct digital synthesizer (DDS) ~\cite{Long2016}. AOM: acousto-optic modulator, EOM: electro-optic modulator, PD: photodetector, Sig Gen: RF signal generator. \textbf{c.} Energy diagram which includes the counterpropagating coupling laser and the resonant $\omega_{LO}/2\pi=4.0~\mathrm{GHz}$ RF LO. \textbf{d.} Transmission spectrum of the probe frequency comb showing electromagnetically induced transparency (EIT) with a full width at half maximum of $3.1~\mathrm{MHz}$ recorded in 100~ms. \textbf{e.} At a modulation frequency $\Delta_{IF}=50~\mathrm{kHz}$, sidebands in the power spectral density are visible on either side of the comb tooth located at the center of the EIT (at $0~\mathrm{MHz}$ in \textbf{d.}), while a comb tooth $10~\mathrm{MHz}$ detuned from the center of the EIT shows no sidebands (labeled in \textbf{d.}). The peaks at $\pm 20~\mathrm{kHz}$ are from laser stabilization and are not related to any atomic interaction.}
    \label{fig:systemDiagram}
\end{figure*}


In this paper, we utilize an electro-optic frequency comb in a self-heterodyne configuration~\cite{Long2016,Dixon2023} to probe the D2 transition in cesium (Cs) atoms which have been prepared in a Rydberg state. We demonstrate a novel signal extraction method where we measure the sidebands which are induced on each of the comb teeth by transmission through the vapor cell. This transmittance modulation can be driven by amplitude or frequency modulation of the RF field impinging on the atoms or via a local oscillator (LO) field detuned from the signal RF field by an intermediate frequency ($\Delta_{IF}$) in the kilohertz to megahertz range.

Figure ~\ref{fig:systemDiagram} shows the experimental setup. The probe laser is locked using a stable Fabry-Perot cavity to $351.72247~\mathrm{THz}$ (i.e., $510~\mathrm{MHz}$ blue-shifted from resonance with the $F=4 - F^\prime=5$ hyperfine transition). Detuning of the comb carrier is necessary to avoid pumping of unwanted hyperfine states~\cite{Long2016}. The total power of the probe laser passing through the vapor cell is $790~\mu\mathrm{W}$, with $740~\mu\mathrm{W}$ at the comb carrier frequency and just $50~\mu\mathrm{W}$ distributed over the 400 total comb teeth (200 on the D2 line and 200 unused). This results in just $125~\mathrm{nW}$ per comb tooth and a peak Rabi frequency of $\Omega_p/2\pi \sim 0.30~\mathrm{MHz}$ per comb tooth~\cite{Sibalic2017}. 
The coupling laser power at the vapor cell is $88~\mathrm{mW}$ $(\mathrm{peak}~\Omega_c/2\pi \sim 1.1~\mathrm{MHz})$ and both beams are Gaussian with a full width at half maximum near $1~\mathrm{mm}$.

The optical frequency comb readout allows us to simultaneously record arbitrary RF modulation frequencies imprinted on the individual comb teeth.
With this technique, it is possible to receive modulated RF signals 
across a wide dynamic range and over complex spectral behavior 
without the need for adjustments to the laser wavelengths. 

\begin{figure}[t]
    \centering
    \includegraphics[width=\linewidth]{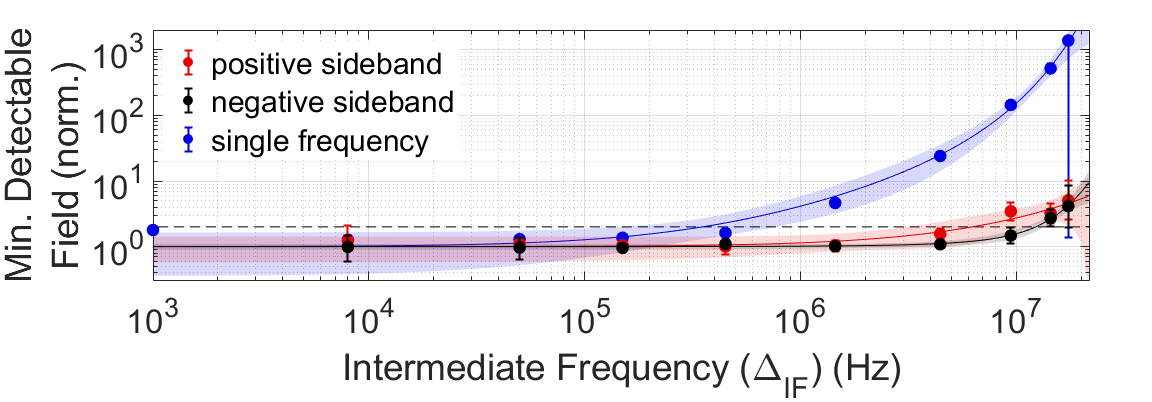}
    \caption{Measured minimum detectable RF field normalized to fit at 0~Hz. Lines report exponential fits and shaded regions give 95\% fit confidence bounds. Error bars report 95\% propagated confidence in measured minimum detectable field. Vertical axis is normalized to the fit at the lowest frequency.}
    \label{fig:sensitivityBandwidth}
\end{figure}

To determine the frequency dependence of the minimum detectable field, we recorded sideband spectra over a wide range of RF powers and $\Delta_{IF}$. Gaussian fits of the sideband spectra were used to extract the signal amplitudes at each setting and the RF signal field equal to the noise floor was recorded. We then fit an exponential to the minimum detectable field and extract the frequency at which the fitted curve crosses the point where the minimum detectable field doubles. We define this as our estimated $3~\mathrm{dB}$ ``sensitivity bandwidth" (see Fig.~\ref{fig:sensitivityBandwidth}), noting its relation to the noise-to-signal ratio roll-off~\cite{Knarr2023}. These fits return sensitivity bandwidths of $6\pm4~\mathrm{MHz}$ and $12\pm1~\mathrm{MHz}$ for the $+$ and $-$~sidebands, respectively. 
We note that these results are obtained with a relatively weak coupling laser and large optical beams. Thus, our measurements suggest a bandwidth improvement originating from the use of an optical frequency comb.

Similar measurements were then performed with a single-frequency probe laser with a power of $19~\mu\mathrm{W}$ ($\Omega_p/2\pi = 3.7~\mathrm{MHz}$). In this case, we find that the $3~\mathrm{dB}$ point occurs near $400~\mathrm{kHz}$. Noting that the bandwidth of the single-frequency measurement could be increased by as much as a factor of 1.5 given a weaker probe laser~\cite{Prajapati2022}, use of a probe frequency comb increases the sensitivity bandwidth by at least an order of magnitude.

With the frequency comb, our measured sensitivity to RF signal field strength is approximately $9~\mathrm{mV/m/\sqrt{Hz}}$ for both sideband datasets. This accounts for the full acquisition time of each measurement, ($100~\mathrm{ms}$) and corresponds to a nominal minimum detectable field of $27~\mathrm{mV/m}$ (at low $\Delta_{IF}$). To compare with typical Rydberg receiver sensitivities, we implement the same technique as Dixon, et. al~\cite{Dixon2023} and extract the central $0.1~\mathrm{MHz}$ portion of the $200~\mathrm{MHz}$ wide sideband spectrum corresponding with the signal peak. This results in an equivalent single frequency sensitivity of $191~\mu\mathrm{V/m/\sqrt{Hz}}$, which is typical of Rydberg atom receivers operating at moderately high principal quantum number.

We note that two separate factors may contribute to the higher bandwidths reported here with the optical frequency comb readout. First, the Rabi frequency per probe comb tooth is very low, and previous studies have shown that bandwidth is inversely proportional to Rabi frequency of the probe laser~\cite{Bohaichuk2022, Prajapati2022, Hu2023}. However, considering the number of comb teeth falling within the EIT linewidth ($\sim8$) results in a much larger Rabi frequency of $2.4~\mathrm{MHz}$. 
Therefore, our observed bandwidth improvement is likely an outcome of the time dynamics between the Rydberg atoms and the $1~\mu\mathrm{s}$ repetition rate comb and may be related to destruction of the dark state. This method has similarities with the theorized bandwidth improvement of up to 100~MHz by spatiotemporal multiplexing probe pulse trains~\cite{Knarr2023}. Future studies of these dynamics and the comb power dependence should lead to further bandwidth and sensitivity improvements, which are expected to help unlock the full potential of Rydberg atom electrometry. 


\textbf{Acknowledgments} Official contribution of the National Institute of Standards and Technology; not subject to copyright in the United States.

\textbf{Disclosures} The authors declare no conflicts of interest.

\textbf{Data availability} The data underlying the results presented in this paper are available at https://doi.org/10.18434/mds2-3159.

\bibliographystyle{plain}

\end{document}